\documentclass[12pt]{article}
\input{epsf.sty}
\def\fo{\hbox{{1}\kern-.25em\hbox{l}}}

\def\slashchar#1{\setbox0=\hbox{$#1$}           % set a box for #1
   \dimen0=\wd0                                 % and get its size 
   \setbox1=\hbox{/} \dimen1=\wd1               % get size of /
   \ifdim\dimen0>\dimen1                        % #1 is bigger 
      \rlap{\hbox to \dimen0{\hfil/\hfil}}      % so center / in box 
      #1                                        % and print #1
   \else                                        % / is bigger 
      \rlap{\hbox to \dimen1{\hfil$#1$\hfil}}   % so center #1
      /                                         % and print /
   \fi}                                         %

\def\hide#1{[hidden stuff]}

\def\beq{\begin{equation}}
\def\eeq{\end{equation}}
\def\eq{\end{equation}}
\def\to{\rightarrow}

\def\mEt{\mbox{${\hbox{$E$\kern-0.6em\lower-.1ex\hbox{/}}}_T$}\, } %missing ET

\def\bsg{\ifmmode B\to X_s\gamma\else $B\to X_s\gamma$\fi}
\def\bsll{\ifmmode B\to X_s\ell^+\ell^-\else $B\to X_s\ell^+\ell^-$\fi}
\def\bstt{\ifmmode B\to X_s\tau^+\tau^-\else $B\to X_s\tau^+\tau^-$\fi}
\def\shat{\ifmmode \hat{s}\else $\hat{s}$\fi}

\newcommand{\newc}{\newcommand}

\newc{\asusy}{\delta a^{\rm SUSY}_\mu}
\newc{\lcal}{\int {\cal L}dt}
 
\newc{\LSP}{{\chi^0_1}}
\newc{\stauR}{{\tilde \tau_R}}
\newc{\stau}{{\tilde \tau_1}}
\newc{\mstop}{m_{\tilde{t}}}
\newc{\mHpm}{m_{H^\pm}}
\newc{\gsim}{\lower.7ex\hbox{$\;\stackrel{\textstyle>}{\sim}\;$}}
\newc{\lsim}{\lower.7ex\hbox{$\;\stackrel{\textstyle<}{\sim}\;$}}
\newc{\ie}{{\it i.e.}}          
\newc{\etal}{{\it et al.}}
\newc{\etc}{{\it etc.}}
\newc{\eg}{{\it e.g.}}          
\newc{\kev}{\hbox{\rm\,keV}}            
\newc{\mev}{\hbox{\rm\,MeV}}            
\newc{\gev}{\hbox{\rm\,GeV}}            
\newc{\tev}{\hbox{\rm\,TeV}}
\newc{\xpb}{\hbox{\rm\, pb}}
\newc{\xfb}{\hbox{\rm\, fb}}

\newc{\mtop}{m_t}
\newc{\mbot}{m_b}
\newc{\mz}{m_Z}
\newc{\mw}{M_W}
\newc{\alphasmz}{\alpha_s(m_Z^2)}
\newc{\swsq}{\sin^2\theta_W}
\newc{\tw}{\tan\theta_W}
\newc{\cw}{\cos\theta_W}
\newc{\sw}{\sin\theta_W}
\newc{\BR}{\hbox{\rm BR}}
\newc{\zbb}{Z\to b\bar}
\newc{\Gb}{\Gamma (Z\to b\bar b)}
\newc{\Gh}{\Gamma (Z\to \hbox{\rm hadrons})}
\newc{\rbsm}{R_b^\hbox{\rm sm}}
\newc{\rbsusy}{R_b^\hbox{\rm susy}}
\newc{\drb}{\delta R_b}

\newc{\sgn}{\mbox{sgn}}
\newc{\tbeta}{\tan\beta}
\newc{\uL}{{\tilde u_L}}
\newc{\uR}{{\tilde u_R}}
\newc{\cL}{{\tilde c_L}}
\newc{\cR}{{\tilde c_R}}
\newc{\tL}{{\tilde t_L}}
\newc{\tR}{{\tilde t_R}}
\newc{\dL}{{\tilde d_L}}
\newc{\dR}{{\tilde d_R}}
\newc{\sL}{{\tilde s_L}}
\newc{\sR}{{\tilde s_R}}
\newc{\bL}{{\tilde b_L}}
\newc{\bR}{{\tilde b_R}}
\newc{\eL}{{\tilde e_L}}
\newc{\eR}{{\tilde e_R}}
\newc{\mhp}{m_{H^\pm}}
\newc{\mhalf}{m_{1/2}}
\newc{\emt}{{e/\mu /\tau}}

\newc{\lR}{\tilde{l}_R}
\newc{\lL}{\tilde{l}_L}
\newc{\nL}{\tilde{\nu}_L}
\newc{\na}{\chi^0_1}
\newc{\nb}{\chi^0_2}
\newc{\nc}{\chi^0_3}
\newc{\nd}{\chi^0_4}
\newc{\ca}{\chi^{\pm}_1}
\newc{\cb}{\chi^{\pm}_2}
\newc{\camp}{\chi^\mp_1}
\newc{\cbmp}{\chi^\mp_1}
\newc{\capos}{\chi^{+}_1}
\newc{\caneg}{\chi^{-}_1}
\newc{\phit}{\phi_t}
\newc{\phib}{\varphi_b}
\newc{\phiew}{\phi_{ew}}
\newc{\htz}{h^0_t}
\newc{\hbz}{h^0_b}
\newc{\hewz}{h^0_{ew}}
\newc{\hsmz}{h^0_{sm}}
\newc{\huz}{h^0_u}
\newc{\hsusyz}{h^0_{susy}}

\newc{\C}{{\cal C}}

\newcommand{\drawsquare}[2]{\hbox{%
\rule{#2pt}{#1pt}\hskip-#2pt%  left vertical
\rule{#1pt}{#2pt}\hskip-#1pt%  lower horizontal
\rule[#1pt]{#1pt}{#2pt}}\rule[#1pt]{#2pt}{#2pt}\hskip-#2pt%  upper horizontal
\rule{#2pt}{#1pt}}% right vertical

\newc{\Dal}{\drawsquare{7}{0.6}}

\def\dofigs#1#2#3{\centerline{\epsfxsize=#1\epsfbox{#2}%
   \hfil\epsfxsize=#1\epsfbox{#3}}}

\def\beq{\begin{equation}}
\def\eeq{\end{equation}}
\def\bea{\begin{eqnarray}}
\def\eea{\end{eqnarray}}
\catcode`@=11
\long\def\@caption#1[#2]#3{\par\addcontentsline{\csname
  ext@#1\endcsname}{#1}{\protect\numberline{\csname
  the#1\endcsname}{\ignorespaces #2}}\begingroup
    \small
    \@parboxrestore
    \@makecaption{\csname fnum@#1\endcsname}{\ignorespaces #3}\par
  \endgroup}
\catcode`@=12

\begin{document}
%Remember: There Is No Cabal.
\begin{titlepage}

\begin{flushright}
Fermilab-Pub-02/231-T 
\\
MCTP-02-48 \\
\end{flushright}

\huge
%\vspace{0.05in}
%\renewcommand{\thefootnote}{\fnsymbol{footnote}}
\bigskip
\bigskip
\begin{center}
{\Large\bf Super-conservative interpretation of muon $g-2$ results
applied to supersymmetry}
\end{center}

\large

\vspace{.15in}
\begin{center}

Stephen P. Martin$^a$ and James D.~Wells$^b$

\small

\vspace{.1in}
{\it $^{(a)}$Department of Physics, Northern Illinois University,
         DeKalb IL 60115 {\rm and} \\
Fermi National Accelerator Laboratory, PO Box 5000, Batavia IL 60510 \\}
\vspace{0.1cm}
{\it $^{(b)}$Physics Department, MCTP,  
University of Michigan, Ann Arbor MI 48109-1120}

\end{center}
 
\vspace{0.15in}
 
\begin{abstract}

The recent developments in theory and experiment related to the anomalous
magnetic moment of the muon are applied to supersymmetry. We follow a very
cautious course, demanding that the supersymmetric contributions fit
within five standard deviations of the difference between experiment and
the standard model prediction. Arbitrarily small supersymmetric
contributions are then allowed, so no upper bounds on superpartner masses
result. Nevertheless, non-trivial exclusions are found.  We
characterize the substantial region of parameter space ruled out by this
analysis that has not been probed by any previous experiment. We
also discuss some implications of the results for forthcoming collider
experiments.

\end{abstract}

\medskip

\begin{flushleft}
hep-ph/0209309 \\
September 2002
\end{flushleft}

\end{titlepage}

\baselineskip=18pt
\setcounter{footnote}{1}
\setcounter{page}{2}
\setcounter{figure}{0}
\setcounter{table}{0}

The relationship between the half-integral spin $\vec s$ of the muon and 
its magnetic 
moment $\vec \mu$ is 
written by convention as
\beq
\vec \mu =g \frac{e\hbar}{2m_\mu c}\vec s
        \equiv (1+a_\mu)\frac{e\hbar}{m_\mu c}\vec s.
\eeq
The non-zero value of $a_\mu$
accounts for radiative corrections to
the semi-classical relationship.
It is
convenient for us to introduce a different variable directly related to
$a_\mu$:
\beq
\delta_\mu \equiv(a_\mu - 11659000\times 10^{-10})\times 10^{10} .
\eeq
The current world average of the 
experimental measurement~\cite{Brown:2001mg,Bennett:2002jb}
of $\delta_\mu$ is
\beq
\delta_\mu^{\rm exp} = 203\pm 8.
\eeq
%It is expected that the error/uncertainty of this measurement will
%go down by perhaps a factor of two when the new sets of data are analyzed.

The computation of the theoretical prediction
for $\delta_\mu$ in the Standard Model (SM) framework has been an 
impressive
on-going effort by many groups~\cite{theory computation}.  
A thorough analysis of
this has been presented recently by Davier \etal~\cite{Davier:2002dy}.  
They
present two results, depending on whether they use $\tau$ decay
data or $e^+e^-$ collider decay as their primary source for understanding the
hadronic loop corrections of the vacuum polarization diagrams
contributing to the muon magnetic moment. The results are
\bea
\delta_\mu^{\rm SM} & = & 
\left \{ \begin{array}{ll}
169.1 \pm 7.8 & (e^+e^- ~\mbox{ based}); \\
186.3 \pm 7.1 &(\tau ~\mbox{decay based}).
\end{array}\right.
\eea
The systematic uncertainties differ in the two approaches, and
the final results are in mild disagreement.  This prompted Davier \etal~to 
not combine the analyses.

The resulting difference between theory and 
experiment~\cite{Davier:2002dy} is
\bea
\delta_\mu^{\rm exp}-\delta_\mu^{\rm SM} & = &
\left \{ \begin{array}{ll} 
33.9\pm 11.2 &(e^+e^- ~\mbox{ based}); \\
16.7\pm 10.7 &(\tau ~\mbox{decay based}).
\end{array}\right.
\eea
Therefore, the current results indicate a quite tantalizing
$3\sigma$ discrepancy~\cite{Hagiwara:2002ma} if one prefers the $e^+e^-$ data,
or a not-so-tantalizing 
$1.6\sigma$ discrepancy if one prefers the $\tau$ decay data.

While the $\tau$ decay based analysis
does depend on significant theoretical input~\cite{Melnikov:2001uw},
in 
this letter we do 
not attempt to 
argue for one theoretical estimate
over another, but rather wish to demonstrate that the muon $g-2$
measurement is interesting even if we take the most wide and conservative
estimate of the theoretical and experimental uncertainties.

To accomplish this goal, we will take the union of the $5\sigma$ allowed
regions of $\delta_\mu^{\rm exp}-\delta_\mu^{\rm SM}$ for the $e^+e^-$
based approach and the $\tau$ decay based approach, and then declare that
the supersymmetric contribution $\delta_\mu^{\rm susy}$ must fall into 
this range.  Numerically,
this works out to
\beq
\label{susy range}
-36.8 < \delta^{\rm susy}_\mu < 89.9.
\eeq
Note that the parameter space associated with 
$\delta_\mu^{\rm susy}\simeq 0$ 
is in the allowed region.  Extremely heavy superpartners will just
decouple from the muon $g-2$ computation and yield a very small
contribution, and so no upper bounds on superpartners can be stated in
this analysis. 

The immediate question now is whether this large allowed range has any
impact on our view of supersymmetric parameter space given the myriad of
other experiments that have been performed over the years, including
collider physics direct searches for superpartners.  Interestingly, the
answer is yes.  Especially at high $\tan\beta$, there are many
combinations of smuon and chargino masses that are ruled out by this
conservative muon $g-2$ analysis that have not been probed by any other
experiment.  Although it is impossible to succinctly characterize the
complete parameter space that is excluded, below we will give several
illustrations which convey the power of the muon $g-2$ experiment even
under this most conservative approach to it.

Some might argue that taking a $5\sigma$ union for the allowed
supersymmetry contribution is too conservative, and the muon $g-2$
experiment is more constraining than what will result from our analysis.
That may well be, but our goal is to arrive at {\it exclusion} results 
that
no reasonable person would quarrel with. In other words, there may be more
supersymmetry parameter space excluded than we presented here, and perhaps
the data is even telling us that some regions of supersymmetry
space are being selected by the data. However, our emphasis here is that 
there
is no reasonable chance that our declared excluded region can ever be
resurrected by future data or analysis. Given some remaining skepticism
about the SM theory computation (e.g., see
ref.~\cite{Ramsey-Musolf:2002cy} regarding the light-by-light
contribution), we feel our super-conservative approach to the data and theory
is reasonable.

The recent body
of work on supersymmetric corrections to muon $g-2$ is 
extensive~\cite{g-2 early days,g-2 last year}. 
Our conventions for parameters and computation of the supersymmetric
contributions to muon $g-2$ follow the details presented in
ref.~\cite{Martin:2001st}.  Our procedure is to compute the supersymmetric
corrections as a function of the heavier smuon mass and the lighter
chargino mass, under some basic restrictions that either simplify the
presentation and are theoretically motivated, or insure that there is no
conflict with other experiments. In all cases we require the following 
conditions be satisfied for parameters at the
weak scale:
\begin{itemize}

\item All supersymmetry parameters such as $\mu$, $M_2$, \etc~are real
(no CP violation effects, so the possibility~\cite{Feng:2002wf} of an
electric dipole moment does not arise).

\item $|\mu|>M_2$, which is well within the expectations
of minimal supersymmetry breaking schemes, and is typically 
valid for non-minimal
scenarios discussed in the literature.

\item $M_1=0.5 M_2$, which is required in simple gaugino mass unification 
scenarios.

\item The scalar cubic coupling of smuons to the Higgs field 
satisfies $|A_\mu|/m_{\tilde \mu_2} < 3$, in the notation 
of 
\cite{Martin:2001st}, in order to
avoid electric charge-violating vacua.
(The particular value 3 chosen here has only a very mild impact on the 
results.)

\item Smuon masses must be greater than 95~GeV to be
consistent with LEP results~\cite{PDG}.
\end{itemize}

The theoretical assumptions given above subsume a very large class of
theoretical models for supersymmetry breaking. These include, but are not 
limited to, 
flavor-preserving minimal supergravity-inspired (``mSUGRA") models and 
minimal gauge-mediated supersymmetry breaking (GMSB) models. By making
further assumptions, one can relate the magnitude of $\mu$ to the other
parameters by requiring correct electroweak symmetry 
breaking within the confines of a particular model. However, there are 
many ways that these commonly made
assumptions can be evaded by simple model extensions. Therefore, in 
keeping with our conservative approach to the data and the SM
calculation, we prefer to maintain as general a model framework as 
possible. More particular assumptions of course lead to stronger 
exclusions.
Again, the assumptions listed above are employed only to construct wieldy 
illustrations of the exclusions that the muon $g-2$ experiment can impose.

\begin{figure}[t]
\dofigs{3.5in}{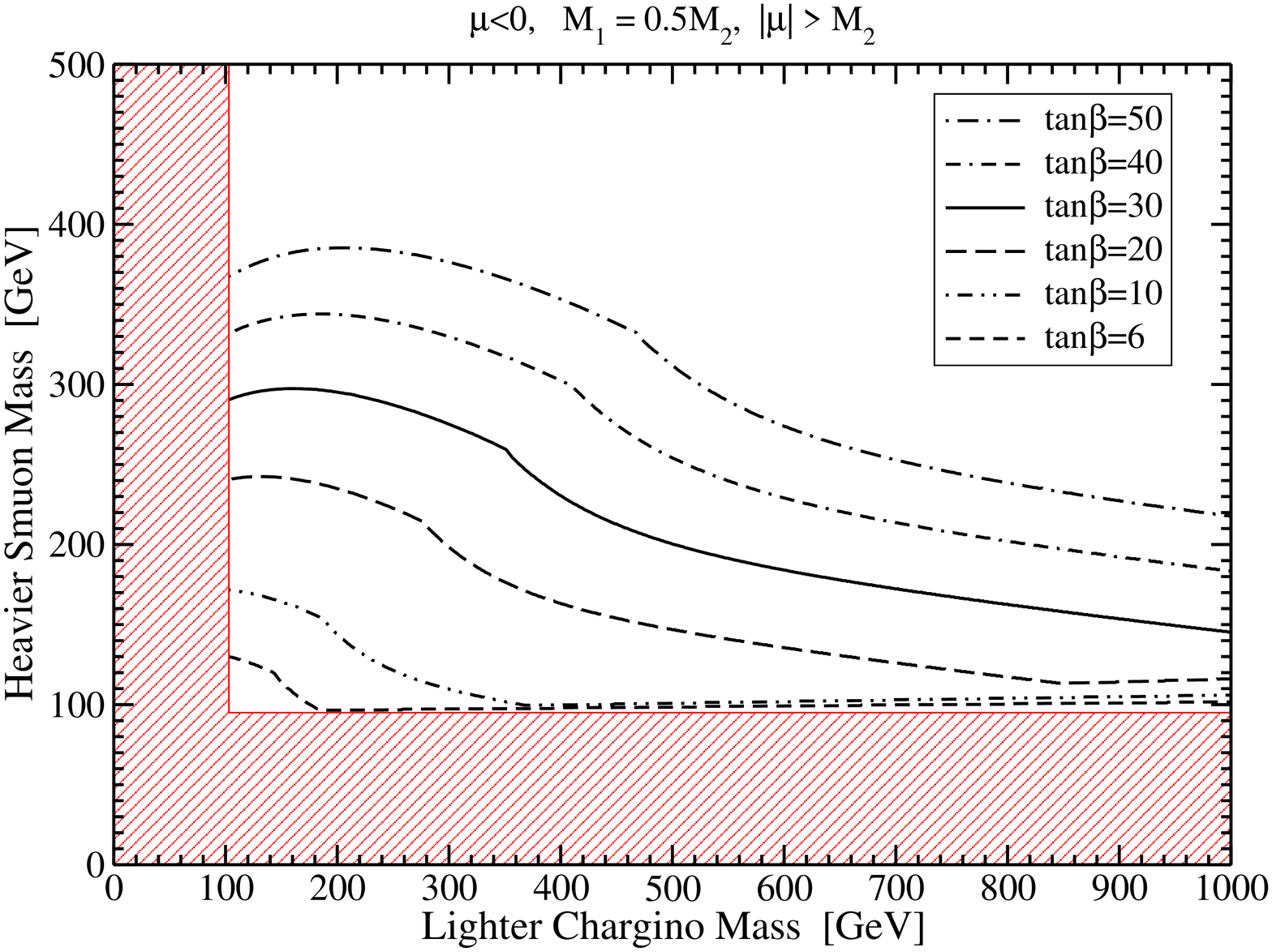}{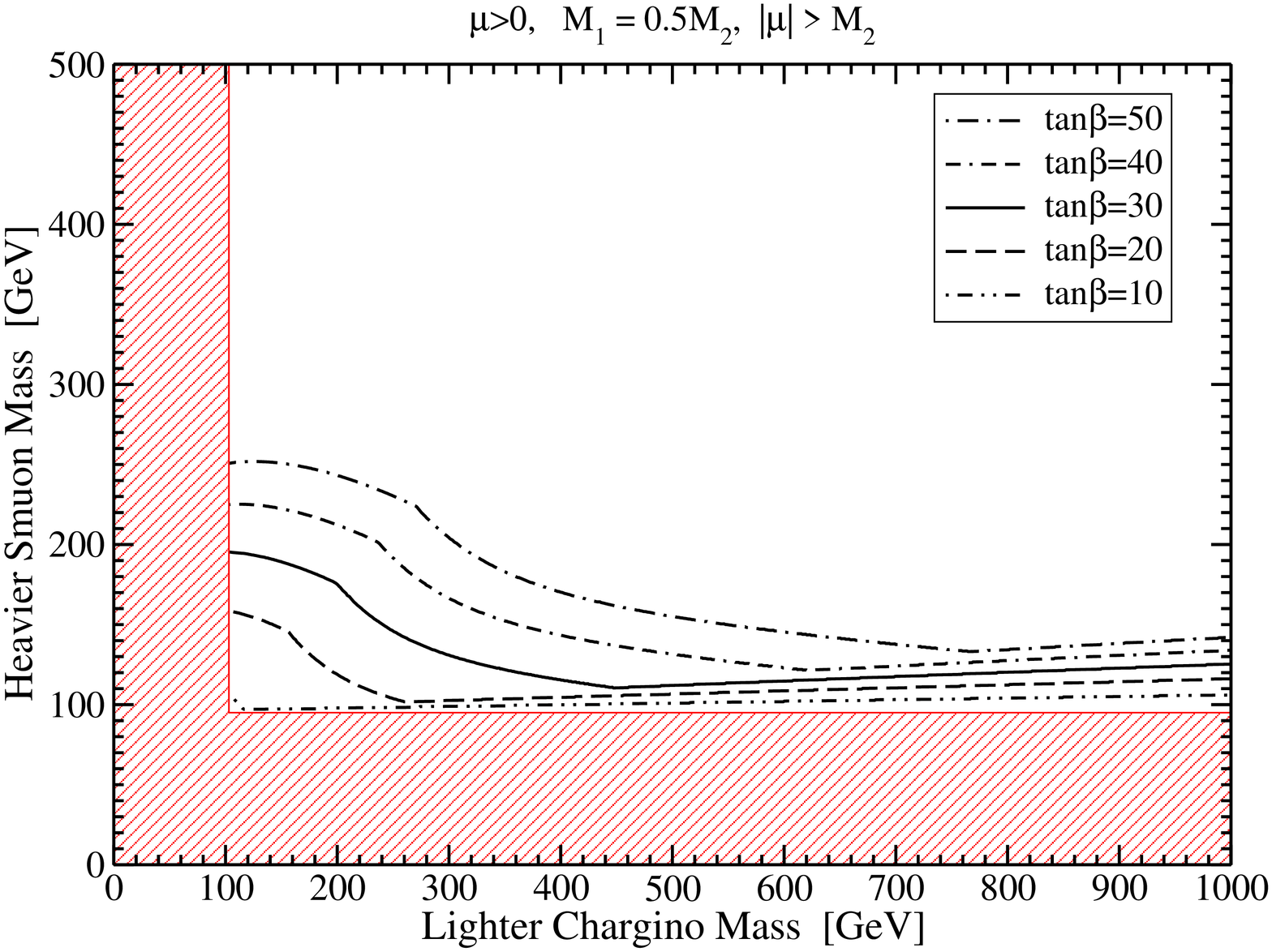}
\caption{The region below each line is excluded by the muon $g-2$
experiment, with the assumptions of gaugino mass unification, $|\mu|>M_2$,
and $m_{\tilde \mu_1}>95\gev$.  The different lines correspond to
different values of $\tan\beta=50,40,30,20,10,6$ from top to bottom. The
left panel is for $\mu<0$ and the right panel is for $\mu>0$. 
The region constrained by direct searches at the LEP collider is shaded.
\label{fig:all}}
\end{figure}

Given our basic assumptions listed above, we show in fig.~\ref{fig:all} 
the excluded area of
the lighter chargino and heavier smuon plane, with different
exclusion contours for various 
values of $\tan\beta$. 
%-------------------------------------------------------------------------
%-------------------------------------------------------------------------
In each case, everything below the exclusion 
contour is inconsistent with the muon $g-2$ 
experiment
and has never before been constrained by another experiment.  We choose
to make the plots using the heavier\footnote{Although we do {\em not}
assume it here, in many theoretical models the heavier smuon is mostly
$\tilde \mu_L$, and plays a more significant role in collider 
phenomenology because of its greater coupling to charginos and 
neutralinos.} smuon mass;
the boundary of the allowed region is then obtained when the other smuon
is not much lighter.
As expected, the excluded region grows 
substantially with larger $\tan\beta$. Even for the relatively
low value of $\tan\beta=6$ there is still a significant portion of
parameter space excluded by the $g-2$ experiment.  Graphically
it looks like a rather small region in the lower left corner
of the graph, but physically it excludes chargino masses as much as 80 GeV
beyond the current limits,  for low smuon masses. This is an important 
part of parameter space for Tevatron searches.
The excluded
region is smaller when we consider $\mu>0$, as this is correlated with
$\delta_\mu>0$, which is less constrained than the $\delta_\mu<0$
($\mu<0$) region.
[The existence of small excluded regions that
stubbornly persist as one looks further along the lighter
chargino mass axis to the right is not related to $g-2$, but rather the
requirement that $m_{\tilde\mu}>95\gev$.  For each
$\tan\beta$, and given a lower limit on $|\mu|>M_2$,
there is a minimum value
of level repulsion induced by the off-diagonal
term of the smuon mass matrix ($m_\mu\mu\tan\beta$) such that it
is impossible to have both $m_{\tilde\mu_1}\simeq 95\gev$ and 
$m_{\tilde\mu_2}\simeq 95\gev$.  Therefore, the
heavier smuon mass must be a few GeV or more above $95\gev$.]

\begin{figure}[t]
\dofigs{3.5in}{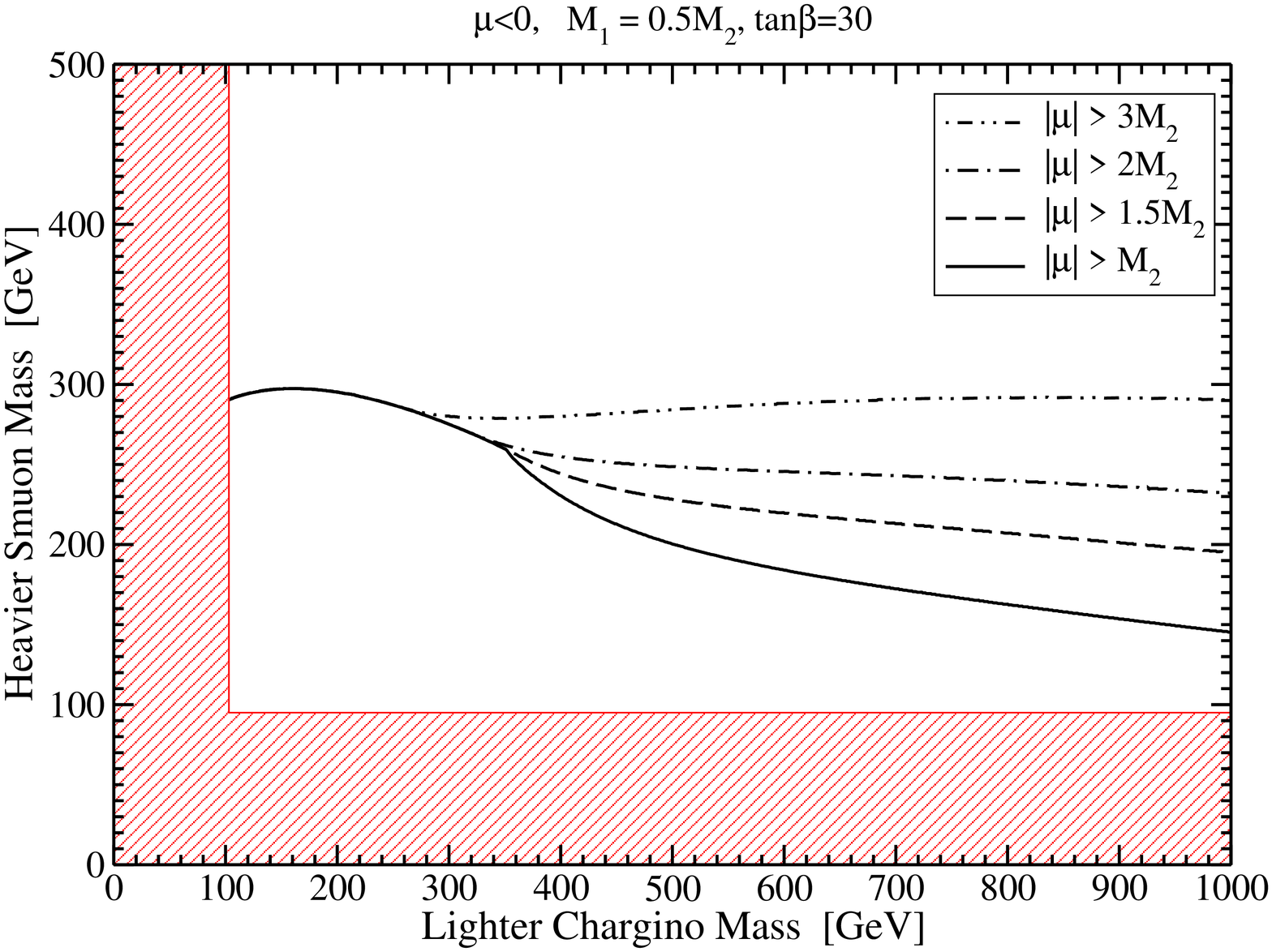}{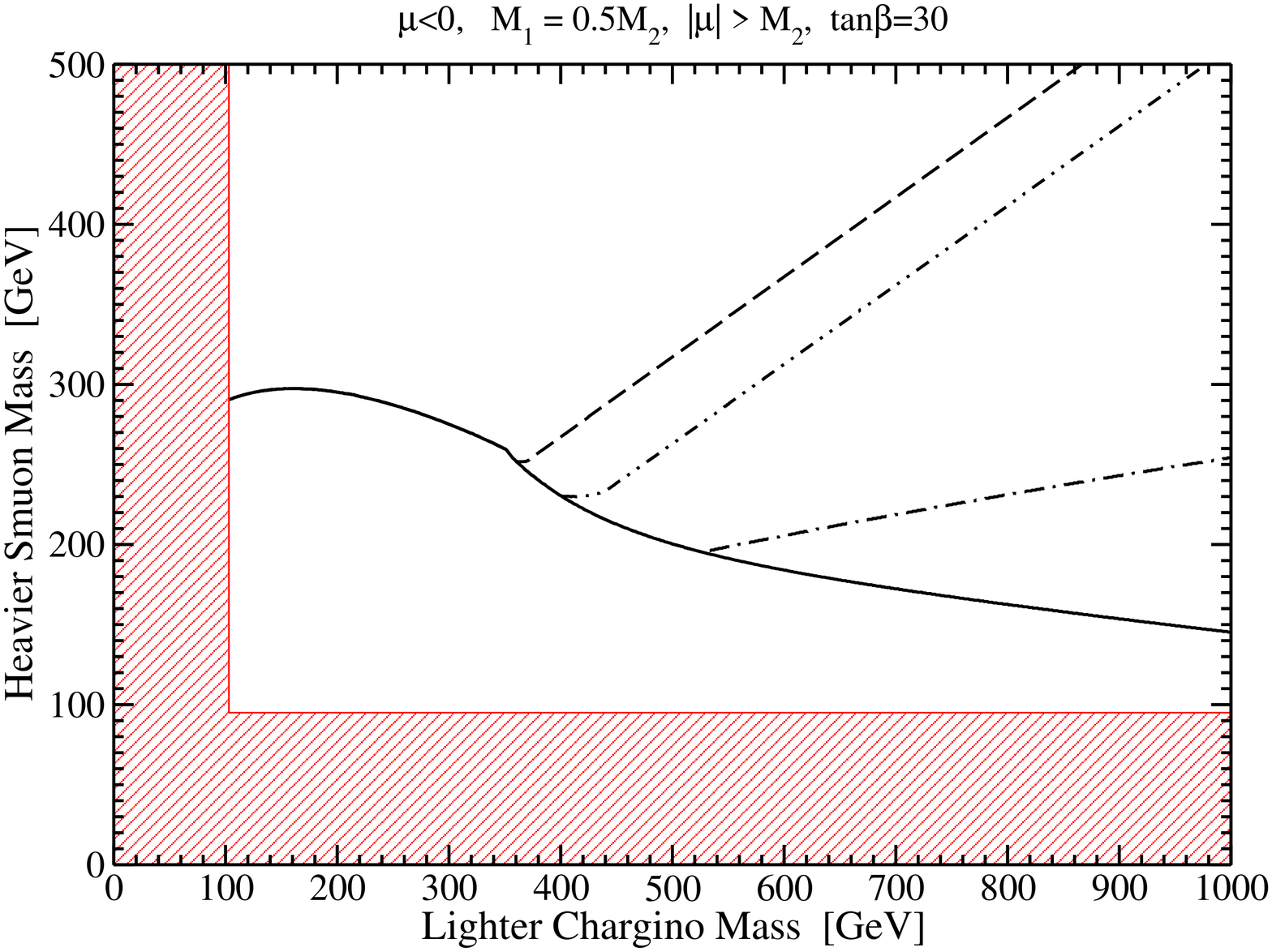}
\caption{The impact of different assumptions is shown, for $\tan\beta=30$ 
and $\mu < 0$. 
The regions under the contour lines are
excluded by the $g-2$ experiment and the stated model assumptions.
In the left panel, we show the excluded regions for 
different choices of the minimum allowed value of $|\mu|/M_2$.
For lighter charginos, the exclusion is set by relatively large $|\mu|$,
so the exclusion contours are not affected.
In the right panel, we show the
excluded areas 
obtained
by adding the requirements (as described in the text) that 
$m_{\tilde\tau_1}>80\gev$ (dash-dotted),
$m_{\chi^0_1}<m_{\tilde\mu_1}$ (dash-dot-dotted), or 
$m_{\chi^0_1}<m_{\tilde\tau_1}$ (dashed).  Notice that these 
additional constraints again have no effect if the chargino is not too 
heavy. 
\label{fig:tb30}}
\end{figure}

Let us now consider how the exclusions depend on the rather conservative 
assumptions we have made. The boundary of the excluded region is
saturated by large (but not arbitrarily large) $|\mu|/M_2$ for lighter 
charginos, and by the minimum
allowed value of $|\mu|$ for sufficiently heavy charginos.
To illustrate this, we show
in the left panel of fig.~\ref{fig:tb30}, for $\tan\beta=30$ and $\mu<0$, 
how the excluded region increases as one raises
the minimum allowed value of $|\mu|/M_2$. (The solid line is the same as
in fig.~\ref{fig:all}.) In many models of supersymmetry breaking, 
$|\mu|/M_2$ is required to be well above 1 in order to have correct 
electroweak symmetry breaking, but we see that increasing
the minimum $|\mu|/M_2$ ratio only affects the 
exclusion contours for quite heavy charginos.
%-------------------------------------------------------------------------
%-------------------------------------------------------------------------

In the right panel of fig.~\ref{fig:tb30}, the dashed, dash-dotted and
dash-dot-dotted lines are stronger exclusion contours that employ
assumptions on the supersymmetric parameter space in addition to the ones
discussed above (with $\tan\beta=30$, $\mu<0$, and $|\mu| > M_2$).  For
the dash-dot line, we add the requirement that $m_{\tilde\tau_1}>80\gev$.  
Implicit in this is an assumption that the diagonal terms of the stau mass
matrix are approximately the same as those of the smuon mass matrix, but the off
diagonal terms are not ($m_\tau\mu\tan\beta$ rather than
$m_\mu\mu\tan\beta$).  In most fundamental theories of supersymmetry
breaking that approximately respect flavor, the diagonal terms of these
mass matrices are the same at a high scale but diverge at the low
scale where the mass eigenvalues need to be evaluated. The mismatch in
diagonal entries between the smuon and stau mass matrices has a smaller
effect than the off-diagonal mismatch on the difference in mass
eigenvalues, so this approximation has more generality than might perhaps
be naively expected.
The excluded region is now bigger 
than that of the solid line,
but the part of the boundary for lighter charginos is still entirely due 
to the muon $g-2$
result. The dashed line further requires that the lightest neutralino is
the lightest superpartner, $m_{\chi^0_1}<m_{\tilde\tau_1}$. This is
motivated by our desire to not have a charged lightest supersymmetric
particle (LSP) in the event that R-parity is exactly conserved and the LSP
is stable.  Significant cosmological constraints apply to charged LSPs and
it is unlikely that a stable, charged LSP can exist without 
modifying the standard scenario
of cosmological history. The dash-dot-dot line in fig.~\ref{fig:tb30}b does
not assume anything about the tau-slepton, and only requires that
$m_{\chi^0_1}<m_{\tilde\mu_1}$.  It is apparent that the additional
constraints just mentioned have no effect on the exclusion curve
when a chargino is light (in
this case, lighter than about 360 GeV).

\begin{figure}[t]
\dofigs{3.5in}{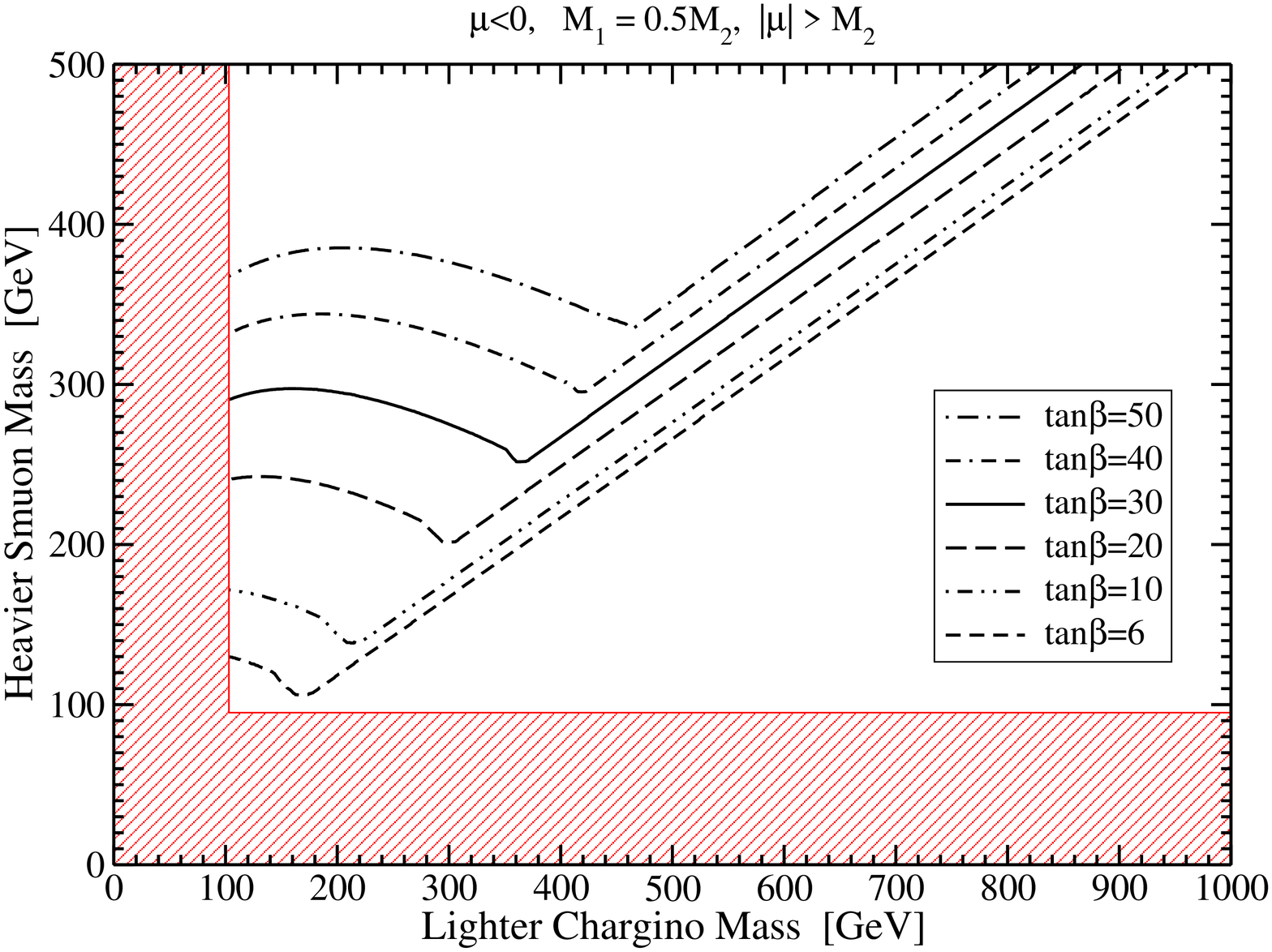}{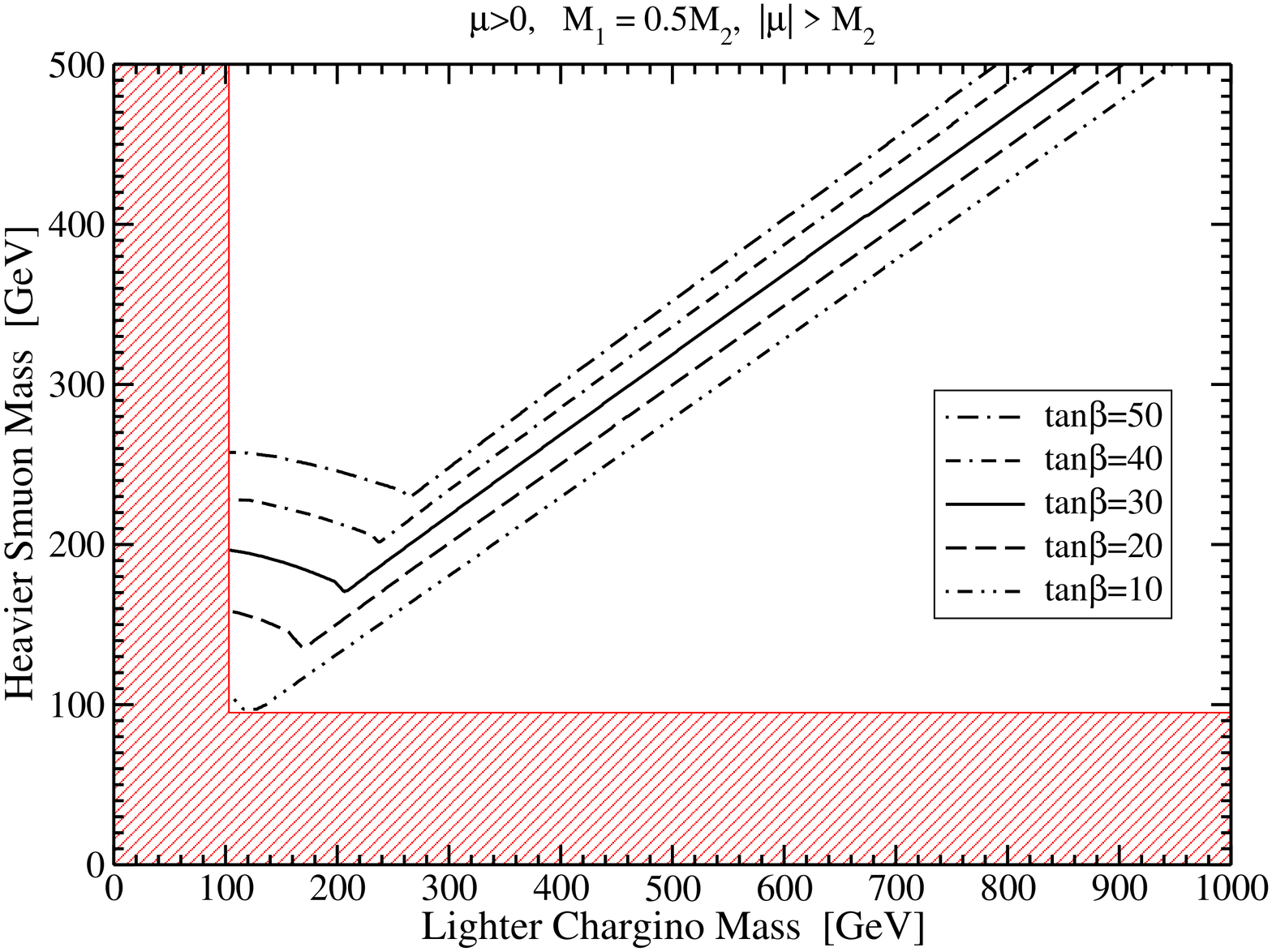}
\caption{As in fig.~\ref{fig:all}, but with the additional constraint
that a neutralino is the lightest supersymmetric particle: $m_{\chi_1^0} < 
m_{\tilde \tau_1}$.
Here again, the regions under the contour lines are
excluded by the $g-2$ experiment and the stated model assumptions.
The left panel is for $\mu<0$ and the right panel is
for $\mu>0$.
\label{fig:con}}
\end{figure}

In fig.~\ref{fig:con} we make the same plots as in fig.~\ref{fig:all}, but
with the additional constraint that the lightest 
supersymmetric particle should be a neutralino ($m_{\chi^0_1} < m_{\tilde 
\tau_1}$).  
The straight parts of the
exclusion contours on the right part of each panel are set by the 
requirement that the
lightest neutralino be the lightest superpartner (less than the stau
mass).  
The remaining part of each exclusion contour curve is due to the muon 
$g-2$
experimental result. The region is significant and demonstrates the
probing/exclusion capability of the muon $g-2$ experiment. Again, even for
the relatively low value of $\tan\beta=6$ there is still a significant
portion of parameter space excluded by the $g-2$ experiment.  

%-------------------------------------------------------------------------
%-------------------------------------------------------------------------

There are several interesting conclusions one can draw from
figs.~\ref{fig:all}, \ref{fig:tb30} and~\ref{fig:con} in addition to just
demonstrating a rather large excluded area.  For example, if $\tan\beta >
30$ and $\mu < 0$ and a chargino is found at or below $360\gev$, there is 
no way that
both smuons could have mass below $250\gev$, given our assumptions.  If we
assume in addition that a neutralino is the lightest supersymmetric
particle, then this lower bound on the heavier smuon mass becomes 
independent of the chargino mass. Heavy slepton mass scales such as this 
would make a
$\sqrt{s} = 500$ GeV linear collider incapable of detecting both sleptons
and measuring their masses and mixings.  Likewise, 
if 
both smuons are
found below about $240\gev$, then a $\sqrt{s} = 500$ GeV linear collider
should be able to discover them and study them with enough luminosity, but
would not be able to discover and study even one chargino
if $\tan\beta > 30$ and $\mu<0$ with our assumptions.

Similar correlating statements apply to the Tevatron and LHC.  For
example, from figs. 66-69 of ref.~\cite{Abel:2000vs} we see that the
Tevatron upgrade will have a significant discovery capability for light
charginos and neutralinos through the trilepton signal. For $\tan\beta$
larger than a few (which is suggested by the failure of LEP to detect a
Higgs scalar boson), the parameter space in which a trilepton signal might
be visible tends to divide into two disconnected regions: a region in
which sleptons are comparable in mass to the charginos and neutralinos,
and a region in which the sleptons are much heavier.

First, consider
the case that sleptons are comparable in mass to the charginos and
neutralinos. The sleptons generally greatly increase the branching
fractions of chargino and neutralino decays into leptons, enhancing the
clean (few backgrounds) trilepton signal. Masses of charginos and
neutralinos up to nearly 200 GeV can be probed in these circumstances.
However, for large $\tan\beta \gsim 10$, this is also the region where the
$g-2$ is very sensitive to supersymmetry. Therefore, for large $\tan\beta$
the future Tevatron trilepton search will have a strong (but not complete)
overlap with the $g-2$ exclusions we have found above, especially for
$\mu<0$. The muon $g-2$ excluded region grows rapidly while the trilepton
search sensitivity tends to fall quickly with increasing $\tan\beta$. So,
in this region of parameter space under our assumptions the present muon 
$g-2$ exclusions take
a very large bite out of the otherwise new territory that the Tevatron
will probe.  The precise maximum values of $\tan\beta$ which the Tevatron
can probe and which are not already ruled out by a conservative
interpretation of the muon $g-2$ results will of course depend strongly on
the Tevatron experimental parameters which are still to be determined.

Considering instead the region of parameter space in which the sleptons
are very massive, the charginos and neutralinos decay branching fractions
to leptons asymptote, respectively, to the values of the $W$ and $Z$ decay
branching fractions to leptons.  This remains true provided the masses are
light enough such that the second neutralino cannot decay into a Higgs
boson plus lightest neutralino.  In the extreme of all scalars decoupling,
the chargino and second lightest neutralino can be discovered or excluded
at the Tevatron with our assumptions if their masses are below about 130 
GeV. In that region,
the Tevatron has no competition from the present muon $g-2$ result and is
entirely complementary to it.

As for the LHC, there are circumstances in which Tevatron and $g-2$ data
would provide significant insights into LHC searches.  For example, if
charginos are found to be light at the Tevatron (e.g., ${\rm mass}\lsim
150\gev$) and $\tan\beta$ is determined to be large (e.g., $\tan\beta
\gsim 30$), the exclusion plots presented here imply that at least one
slepton mass has to be greater than about $290\gev$ ($190\gev$) for $\mu
<0$ ($\mu>0$). Sleptons with mass greater than about 300 GeV would be
very challenging to directly detect at the LHC~\cite{Baer:1993ew}. Or, if 
$\tan\beta$
has not been measured by the time LHC collects data, discovery of light
sleptons and a light chargino with masses less than about 190 GeV would
rule out large $\tan\beta>30$ supersymmetry, for either sign of 
$\mu$, because of its incompatibility
with the muon $g-2$ experimental result.

Our central point is that even the most conservative view of the data
produces a large region of supersymmetry excluded only by the muon $g-2$,
and no other experiment. The implications have the potential to be very
important for future experiments.  As we detailed in our previous
paper~\cite{Martin:2001st}, the constraints from $B(b\to s\gamma)$, Higgs
mass, and relic abundance are not necessarily correlated in any meaningful
way with $g-2$, and so the $g-2$ experiment is firmly established as an
independently powerful probe of supersymmetry.

%%%%%%%%%%%%%%%%%%%%%%%%%%%%%%%%%%%%%%%%%%%%%%%%%%%%%%%%%%%%%%%%%%%%%%%%%
\bigskip \noindent {\it Acknowledgments:}   We thank G. Kane for 
helpful discussions.
SPM is supported in part by
the National Science Foundation grants PHY-9970691 and PHY-0140129
and a State of Illinois Higher Education Cooperation Act grant, and 
JDW
is supported in part by the Department of Energy and the Alfred P. Sloan
Foundation.

%%%%%%%%%%%%%%%%%%%%%%%%%%%%%%%%%%%%%%%%%%%%%%%%%%%%%%%%%%%%%%%%%%%%%%%%

\end{document}